\documentclass[aps,pre,preprint,superscriptaddress,showpacs]{revtex4}

\usepackage{graphicx}

\begin{document}

\title{Complexity transitions in global algorithms for sparse linear
systems over finite fields}

\author{A. Braunstein}
\email{abraunst@ictp.trieste.it}
\affiliation{International Center for Theoretical Physics, Strada
Costiera 11, P.O. Box 586, I-34100 Trieste, Italy}
\affiliation{SISSA, via Beirut 9, I-34100 Trieste, Italy}
\author{M. Leone}
\email{micleone@ictp.trieste.it}
\affiliation{International Center for Theoretical Physics, Strada
Costiera 11, P.O. Box 586, I-34100 Trieste, Italy}
\affiliation{INFM and SISSA, via Beirut 9, I-34100 Trieste, Italy}
\author{F. Ricci-Tersenghi}
\email{Federico.Ricci@roma1.infn.it}
\affiliation{International Center for Theoretical Physics, Strada
Costiera 11, P.O. Box 586, I-34100 Trieste, Italy}
\affiliation{Dipartimento di Fisica, Universit\`a di Roma ``La
Sapienza'', Piazzale Aldo Moro 2, I-00185 Roma (Italy)}
\author{R. Zecchina}
\email{zecchina@ictp.trieste.it}
\affiliation{International Center for Theoretical Physics, Strada
Costiera 11, P.O. Box 586, I-34100 Trieste, Italy}
\affiliation{Laboratoire de Physique Th\'eorique et Mod\`eles
Statistiques, Universit\'e Paris Sud, 91405 Orsay, France}

\date{\today}

\begin{abstract}
We study the computational complexity of a very basic problem, namely
that of finding solutions to a very large set of random linear
equations in a finite Galois Field modulo $q$.  Using tools from
statistical mechanics we are able to identify phase transitions in the
structure of the solution space and to connect them to changes in
performance of a global algorithm, namely Gaussian elimination.
Crossing phase boundaries produces a dramatic increase in memory and
CPU requirements necessary to the algorithms. In turn, this causes the
saturation of the upper bounds for the running time.  We illustrate
the results on the specific problem of integer factorization, which is
of central interest for deciphering messages encrypted with the RSA
cryptosystem.

\end{abstract}


\maketitle

\section{Introduction}

The methods and concepts of statistical physics of disordered systems
constitute a very useful tool for the understanding of the onset of
computational complexity in randomly generated hard combinatorial
problems.  Once the optimization problems are translated into zero
temperature spin glass problems, one may study the geometrical changes
in the space of solutions as symmetry breaking phenomena.  In this
context one may view the exponential regimes of randomized search
algorithms as out-of-equilibrium phases of stochastic processes.

However, combinatorial problems are not always exponentially hard:
Problems that can be solved in polynomial time, even in their
worst-case realizations compose the so called Polynomial (P) class
\cite{GAREY}.  Such problems are often of great practical relevance
and are tackled using large scale computations.  Examples can be found
in all disciplines: In physics, just to make one example, one may
study ground states of 2D spin glass like Hamiltonians resorting to a
polynomial max-cut algorithm \cite{ALAVA}.  The major application are
obviously found in engineering: Examples are design problems (finite
elements methods), control theory (convex optimization), coding theory
(parity check equations) and cryptography (integer factorization).

Due to the practical relevance of the problems and to the typically
large number of variables used for their encoding, that is the size of
the problems, it is of basic interest to look at the fine structure of
the class P in order to concretely optimize the computational
strategies.  For instance, in error correcting codes it is crucial to
have algorithms that converge in {\em linear} time with respect to the
number of encoded bits, any power larger than one being considered of
no practical interest.

Quite in general, the trade-off between time and memory resources is
the guiding criterion which selects the algorithms used in real-world
applications.  Roughly speaking polynomial algorithms can be divided
in different groups depending on the solving strategy they
implement. The main groups are local algorithms (e.g.\ greedy/gradient
methods), global algorithms (e.g.\ Gaussian elimination or Fourier
transforms methods), iterative algorithms (e.g.\ Lanczos method) and
parallel algorithms.  See Ref.~\cite{CLR} for a basic introduction to
the subject.

In what follows we shall study a prototype problem of the P class,
that is the problem of solving large and random sparse systems in some
Galois field GF($q$).  Working in GF($q$) is completely equivalent to
perform any operation modulo $q$.

Firstly, we give a precise analysis of the computational features for
non-trivial ensembles of random instances. By a statistical mechanics
study, we look into the -- symmetry breaking -- geometrical structure
of the space of solution thereby providing an explanation for the
changes in the power law behavior observed in different algorithms.
Moreover, we are able to predict and explain in terms of clustering of
solutions, the memory catastrophe found in global algorithms such as
Gaussian elimination.  Such an effect seriously hampers application of
this sort of global algorithms in many circumstances, one example
being symbolic manipulations.  This memory catastrophe induce in turn
an even more dramatic increase in CPU time, which make large problems
unaffordable above the dynamical threshold $\gamma_d$ (see below for
its definition).

Secondly, we consider a specific ``real-world'' application, namely
the Integer Factorization problem used in RSA public key cryptography
\cite{RSA}. By a non-trivial mapping of the factoring problem on a
sparse linear system modulo 2, endowed with a quite peculiar
statistical distribution of matrix elements, we analyze which are the
characteristic geometrical properties of solutions that are
responsible for the usage of specific algorithms and constitute the
possible bottleneck for the near future.

Interestingly enough, the changes in both time or memory requirements
during the solution process of sparse systems can be interpreted in
physical terms as a dynamical transition at which the phase space of
the associated physical systems becomes split into an exponential
number of ergodic components.  While it is to be expected that local
algorithms get stuck by local minima at such phase boundary, it is
less obvious to predict which is the counterpart of the dynamical
transition in global algorithms, for which polynomial time convergence
is guaranteed even for the hardest instances. Indeed the dynamical
transition manifests itself as a phase transition in the computational
requirements which in turn leads to a slowing down phenomenon that
saturates the upper bound for the convergence time. Such a change of
scale in memory requirements constitute a serious problem for hardware
implementations of large scale simulations.

Hopefully, the paper is written in a style accessible to a cross
disciplinary audience.

\section{Random Linear systems in GF(2): rigorous results and
statistical mechanics analysis}

The theory of random equations in finite fields is shared by
probability, combinatorics and algebra \cite{KOLCHIN}.

For the sake of simplicity we limit our statistical mechanics analysis
to GF(2) rather than GF($q$), the extension to $q>2$ being
straightforward though technically involved.

As is well known in the context of error correcting codes
\cite{Sourlas}, solving a sparse linear system modulo 2 is equivalent
to finding the zero temperature ground states of a class of multiple
degree interactions $p$-spin models on diluted random graphs.

Let us consider a random linear system in GF(2) in the form $\hat{A}
\vec{x} = \vec{y} \;\; \text{mod} \; 2$, where $\hat{A}$ is a 0-1
matrix of dimension $M \times N$. For each of its specific choices
$\hat{A}$ can be interpreted as the contact matrix of a particular
random (hyper-)graph belonging to a specific ensemble. The class of
random matrices we shall deal with are defined by the fraction of rows
$a_k$ with $k$ non zero elements. The latter are placed uniformly at
random within each row.

We focus on matrices that lead to graphs with an average connectivity
value $\langle k \rangle = \sum_k a_k k$ finite and much less than
both $M$ and $N$. We are interested to the limit of very large
matrices, where we can assume $N,M \to \infty$ with a finite ratio
$\gamma \equiv M/N$.

This is the regime in which a study of the computational cost is
important in that it applies directly to large scale computations. In
the limit $N,M \to \infty$ average quantities characterizing the
system (e.g.\ the average fraction of violated equations) are known to
be equal to the most probable values (i.e.\ their probability
distribution is strongly peaked \cite{self-average}) and therefore
single random large systems behave as the average over the ensemble.

We will always assume $a_1=0$ at the beginning, since rows with a
single one corresponds to trivial equations which can be removed a
priori from the set.

The equivalence between linear systems and spin models is quite
straightforward.  We start from a set of linear equations in GF(2),
$\hat{A} \vec{x} = \vec{y}$, and we build up a spin Hamiltonian whose
ground state energy $E_{gs}$ counts the minimal number of unsatisfied
equations.  In the case where $E_{gs}=0$, ground state configurations
will correspond to solutions of the original set of linear equations
and the zero-temperature entropy will count the number of such
solutions.

The construction is done as follows: For every equation, labelled by
$i \in [1 \dots M]$, let us define the set of variables $\vec{x}$
entering equation $i$ as
\begin{equation}
v(i) \equiv \{ j \in [1 \dots N] : A_{ij}=1 \} \quad .
\end{equation}
With the transformation $s_j=(-1)^{x_j}$ and $J_i=(-1)^{y_i}$, we have
that every equation can be converted in a term of the Hamiltonian
through
\begin{equation}
\sum_{j=1}^N A_{ij} x_j = y_i \Leftrightarrow \sum_{j \in v(i)} x_j =
y_i \Leftrightarrow \prod_{j \in v(i)} s_j = J_i \quad ,
\end{equation}
where the multi-spin interaction contain at least 2 spins since we set
$a_1=0$.  Then the Hamiltonian
\begin{equation}
H = \frac12 \left[ M - \sum_{i=1}^M J_i \prod_{j \in v(i)} s_j \right]
\quad ,
\end{equation}
fits the above requirements and can be used in the analytical
treatment.

A better form for the above Hamiltonian can be obtained grouping
together $k$-spin terms with the same $k$, that is
\begin{equation}
H = \frac12 \left[ M - \sum_k \sum_{i_1 < i_2 < \dots < i_k} J_{i_1
i_2 \dots i_k} s_{i_1} \dots s_{i_k} \right] \quad ,
\label{eq:ham}
\end{equation}
where $s_i=\pm1$ are Ising spins and the couplings $J_{i_1 i_2 \dots
i_k}$ are i.i.d.\ quenched random variables taking values in
$\{0,\pm1\}$.  The total number of interactions, that is of terms with
$J \neq 0$, is M, and the energy is zero if and only if all the
interactions are satisfied.  For each unsatisfied interaction the
energy increases by 1.

The fraction of interactions of $k$-spin kind is $a_k$ and thus the
probability of having $J_{i_1 i_2 \dots i_k} \neq 0$ equals $a_k M /
{N \choose k} \simeq \gamma a_k k! / N^{k-1}$, while the sign of
$J_{i_1 i_2 \dots i_k}$ depends on the probability distribution of the
components of $\vec{y}$,
\begin{equation}
P(J_{i_1 i_2 \dots i_k}) = \left[ 1 - \frac{\gamma a_k k!}{N^{k-1}}
\right] \delta(J_{i_1 i_2 \dots i_k}) + \frac{\gamma a_k k!}{N^{k-1}}
\Big[ p\; \delta(J_{i_1 i_2 \dots i_k} - 1) + (1-p)\, \delta(J_{i_1
i_2 \dots i_k} + 1) \Big] \quad ,
\end{equation}
where $p \in [0,1]$ controls the fraction of zeros in $\vec{y}$.  As
long as the system admits at least one solution, it can always be
brought by a gauge transformation in the form with $p = 1
\Leftrightarrow \vec{y} = \vec{0}$.  This corresponds to have positive
or null couplings only, like in a diluted ferromagnetic model.

In order to make a connection between the behavior of solving
algorithms and the structure of the matrix $\hat{A}$, we study the
geometrical properties of the space of solution, i.e.\ ground states
of (\ref{eq:ham}), as a function of $\gamma$ for non-trivial choices
of $\{a_k\}$.  We may have access to the structure of such a space by
just performing the $T=0$ statistical mechanics analysis of the spin
glass model, with control parameter $\gamma$.

For $\gamma$ large enough, at say $\gamma_c$, the system of equations
becomes over-determined and some of the equations can no longer be
satisfied.  This fact is reflected in the ground state energy of the
associated spin glass model becoming positive.  The interesting aspect
of the problem is that, under proper conditions, there appears a
clustering phenomenon with macroscopic algorithmic consequences at
some intermediate value $0 < \gamma = \gamma_d < \gamma_c$.  We will
focus our attention on the latter transition, thus assuming a priori
that at least one solution always exist.  This allow us to fix
$\vec{y} \equiv \vec{0}$ hereafter.

The complete picture of the typical structure of the solution space
can be obtained through a replica calculation, following a well tested
scheme in diluted systems, as in \cite{RIWEZE} and \cite{LERIZE}.  The
results of such a calculations have been recently confirmed by a
rigorous mathematical derivation \cite{MERIZE,CDMM}.  We avoid here to
repeat standard calculations already presented in \cite{RIWEZE,LERIZE}
and extensively reviewed in \cite{LEONE}, and we directly present the
results.

Due to the zero energy condition ($E_{gs}=0$ for $\gamma < \gamma_c$),
the dominance of thermodynamical states is purely to be determined in
entropic terms.  Defining $S_0(\gamma)$ as the logarithm of the number
of solutions to $\hat{A} \vec{x} = \vec{0}$ divided by $N$, we have
that
\begin{equation}
S_0(\gamma) = S(m,\gamma) = \log(2) \left[ (1-m) [1 - \log(1-m)]
-\gamma \sum_{k \ge 2} a_k (1 - m^k) \right] \quad ,
\label{entropy}
\end{equation}
where $m$ solves
\begin{equation}
G(m) = 1 - m - e^{-\gamma \sum_{k \ge 2} k a_k m^{k-1}} = 0 \quad .
\label{mag}
\end{equation}
When more than one solution to Eq.(\ref{mag}) exist, the one
maximazing $S(m,\gamma)$ must be chosen.

At fixed $\{a_k\}$, one can study the phase diagram as a function of
$\gamma$.  At low enough $\gamma$, Eq.~(\ref{mag}) has only the
trivial solution $m=0$ and the system is paramagnetic with entropy
$S(0,\gamma)=\log(2)\;(1-\gamma)$. Typically a non trivial magnetized
solution for the order parameter, $m^*>0$, appears at a value
$\gamma_d$ such that
\begin{equation}
G(m^*)=0 \text{ and } \left. \frac{\partial G(m)}{\partial m}
\right|_{m=m^*} = 0 \quad .
\label{gammadin}
\end{equation}
This solution becomes entropically favored at a value $\gamma_c$ found
solving
\begin{equation}
S(0,\gamma_c) = S(m^*,\gamma_c) \quad .
\end{equation}

The crucial observation is the following. At $\gamma_d$, together with
the magnetized solution, there appear other spin glass solutions to
the saddle-point equation.  In particular, it can be shown
\cite{FLRZ,MERIZE} that the difference between the paramagnetic and
the ferromagnetic entropies,
\begin{equation}
\Sigma(\gamma) = S(0,\gamma) - S(m^*,\gamma) \quad ,
\end{equation}
gives the configurational entropy of the problem, that is the number
of clusters of solutions~\footnote{Two solutions belong to the same
cluster (resp.\ to different clusters) if their Hamming distance is
$\mathcal{O}(1)$ [resp.\ $\mathcal{O}(N)$].}. There exist
$\exp[\Sigma(\gamma) N]$ well separated clusters [Hamming distances
$\sim \mathcal{O}(N)$], each one containing a number
$\exp[S(m^*,\gamma) N]$ of closed solutions [Hamming distances $\sim
\mathcal{O}(1)$].

This clusterizations has two main consequences.  Local algorithms for
finding solutions running in linear time in $N$ stop converging
\cite{RIWEZE}: this is the typical situation for greedy algorithm
which get stuck in one of the most numerous local minima at a positive
energy.

Global algorithms, which are guaranteed to converge in polynomial
time, need to keep track along computation of this complex structure
of solutions and a memory linear in $N$ turns out to be insufficient,
as we will show below.

The simple cases $\{ a_2=1 ; a_{k \neq 2}=0 \}$ and $\{ a_3=1 ; a_{k
\neq 3}=0 \}$ are particularly illuminating \cite{LERIZE}.  Self
consistency equations for the order parameter and for the ground state
entropy can be immediately retrieved from the general formulas
(\ref{mag}) and (\ref{entropy}).

The first case represents a simple Ising spin model on a random
graph. The analysis of thermodynamic phases shows a trivial
paramagnetic region at low $\gamma$, followed by a second order phase
transition at $\gamma_d = \gamma_c = 1/2$ where both the ground state
energy and the magnetization become positive in a continuous way.
From the linear system point of view, working with sparse equations
with only 2 variables per row is always easy, i.e.\ algorithms have no
slowing down. The solving process progressively fixes variables and
smoothly goes on until a complete solution has been found. Indeed,
fixing one variable immediately fixes the other one within the
equation, and this goes on in a cascade process that prevents the
accumulation of too long symbolic memory-taking expressions.

The second case of the 3-spin model was tackled in full detail in
\cite{RIWEZE,FLRZ}.  It shares the general characteristics of all
models without, or at most with a very small fraction of, 2-spin
terms. In this case a gap between $\gamma_d$ and $\gamma_c$ opens up
and a non vanishing configurational entropy $\Sigma$ appears there.
At $\gamma_c$ the magnetization jumps to a finite value.

For a general choice of $\{a_k\}$, the configurational entropy reads
\begin{equation}
\Sigma(\gamma) = \log(2) \left[ 1 - (1-m) [1 - \log (1-m)] + \gamma
\sum_{k \ge 2} a_k m^k \right] \quad ,
\label{compl}
\end{equation}
where $m$ is the largest solution to Eq.~(\ref{mag}).  As discussed in
Ref.~\cite{FLRZ}, the values of $\gamma_d$ and $\gamma_c$ are found as
the points where $\Sigma(\gamma)$ first appears with a non zero value
and where it reaches zero again.

The algorithmic consequences of having $\Sigma(\gamma) > 0$ have been
already exposed in Refs.~\cite{RIWEZE,FLRZ}: For $\gamma > \gamma_d$ a
glassy state with positive energy arises, which traps any local
dynamics, preventing it to converge towards the ground state of zero
energy.  We conjecture the counterpart on global algorithms, such as
Gaussian elimination, to be that the resolution time increases with
$N$ faster than linear.

In the next section we will check the above conjecture with two
different Gaussian elimination algorithms, none of which is able to
solve the system in linear time for $\gamma > \gamma_d$.

\section{Algorithms behaviour}

In this section we analyze the performances of a couple of different
`Gaussian elimination' algorithms, their difference being in the order
equations are solved.  We will measure the number of operations and
the size of the memory required for the solution of a set of linear
equations, that is the complexity for finding all solutions to $\hat{A}
\vec{x} = \vec{y}$.

We will see that, for a generic ensemble of random problems, any
algorithm undergoes an easy/hard transition at a certain $\gamma$
value, which can not be pushed beyond the dynamical transition
threshold $\gamma_d$.  In this context we call {\em easy} such
problems which are solvable with a CPU-time and memory of order $N$,
and {\em hard} those requiring resources scaling with $N^\alpha$,
where $\alpha>1$.

Given a set of $M$ linear equations in $N$ variables, Gaussian
elimination proceeds as follows [for concreteness we will always work
in GF(2)]: At each step, it takes an equation, e.g.\
$x_1+x_2+x_3=y_1$, solves it with respect to a variable, e.g.\
$x_1=x_2+x_3+y_1$, and then it substitutes variable $x_1$ with the
expression $x_2+x_3+y_1$ in all the equations still unsolved.  This
procedure gives all the solutions to any set of linear equations in,
at most, $\mathcal{O}(N^3)$ steps and using $\mathcal{O}(N^2)$ memory.
Nevertheless this bounds only holds in the worst case, namely when the
matrix $\hat{A}$ is dense.  Very often, in actual applications, the matrix
is sparse and the algorithm is faster.  We define sparse a matrix with
$\mathcal{O}(N)$ ones and dense that with $\mathcal{O}(N^2)$ ones.

In order to analyze the computational complexity of this problem, and
its connections to phase transitions, we focus on a specific ensemble
of random problems, generalizations to other ensembles being
straightforward.  We choose sets of $M=\gamma N$ linear equations,
each one containing exactly $k=3$ of the $N$ variables, taking values
in GF(2).  Thus the connectivity of a variable, defined as the number
of equations this variable enters in, takes values from a Poissonian
distribution of mean $3 \gamma$.

For very large $N$, that is in the thermodynamical limit, we are
interested in how the complexity changes with $\gamma$.  Moreover, for
a fixed $\gamma$ such that the problem is hard, we would like to know
when (in terms of the running-time $t$) and why the algorithm becomes
slower and slower.

\begin{figure}
\includegraphics[width=0.7\columnwidth]{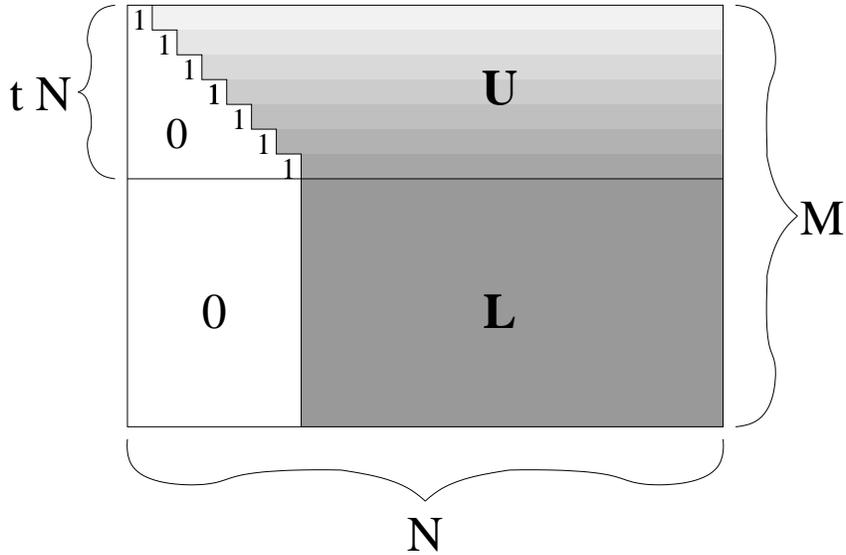}
\caption{Typical shape of the $\hat{A}_t$ matrix after $tN$ steps of
Gaussian elimination.}
\label{matrix}
\end{figure}

\begin{figure}
\includegraphics[width=\columnwidth]{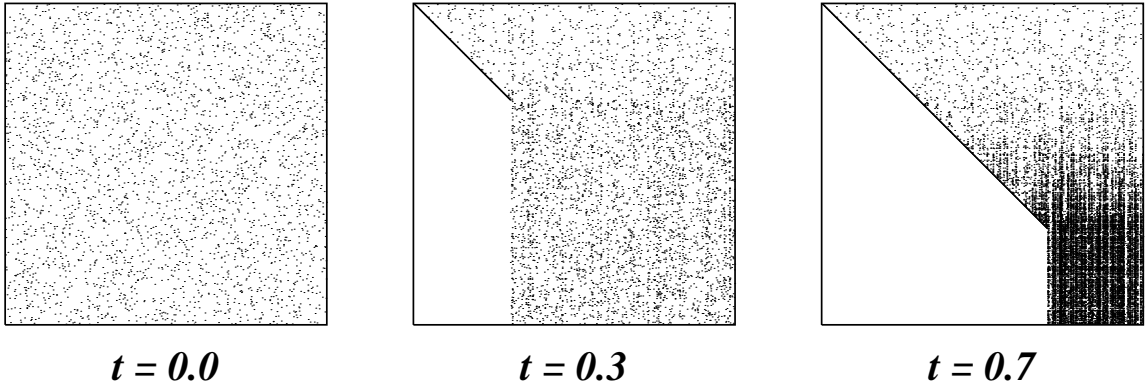}
\caption{The evolution of the $\hat{A}_t$ matrix for a specific $1024
\times 1024$ random system. Every dot corresponds to a 1 entry.}
\label{matrice}
\end{figure}

The running-time $t$ is measured as the number of equations already
solved, normalized by $N$, and thus takes values in $[0,\gamma]$.
$\hat{A}_t$ is the matrix representing the set of equations after $tN$
steps, and it has the form shown in Fig.~\ref{matrix}.  See
Fig.~\ref{matrice} for the actual shape of $\hat{A}_t$ in a specific case
with 1024 equations in 1024 variables.  For ease of simplicity, we
have reordered the variables and the equations of the system, such
that, at the $i$-th step, we solve the $i$-th equation with respect to
$x_i$.  With this choice the left part of the matrix $\hat{A}_t$ has ones on
the diagonal and zeros below.  The right part can be naturally divided
in an upper part $U$ and a lower one $L$.  The density of ones in the
$L$ matrix --- let us call it $\rho(t;\gamma)$ --- is uniform and
depends on the initial $\gamma$, the time $t$ and the algorithm used
for solving the linear system.  The density of ones in the $U$ part is
not uniform and varies from row to row, as shown in Fig.~\ref{matrix}
with gray tones.  For continuity reasons the density at the $m$-th row
of $U$ is exactly $\rho(m/N;\gamma)$.  Then $U$ is sparse or dense
depending on whether $L$ is.  Defining $k(t;\gamma) = \rho(t;\gamma) N
(1-t)$ the average number of ones per row in $L$, we have that a
sparse (resp.\ dense) matrix corresponds to having a finite $k$
(resp.\ $\rho$).

At each time step, the number of operations required are directly
related to the density of the matrix $\hat{A}_t$ and thus to that of $L$.
More specifically, solving with respect to the variable in the upper
left corner of $L$, the number of operations is proportional to the
number of ones in the first row of $L$, i.e.\ $k(t;\gamma)$, times the
number of rows of $L$ having a one in the first column, i.e.\
$\rho(t;\gamma) N (\gamma-t)$, and thus equals
\begin{equation}
k(t;\gamma) \rho(t;\gamma) N (\gamma-t) = k^2 \frac{\gamma-t}{1-t} =
N^2 \rho^2 (\gamma-t) (1-t) \quad .
\end{equation}
Then, if the matrix $L$ is sparse a finite number of operations per
step is enough, while $\mathcal{O}(N^2)$ operations are required when
$L$ is dense.  Integrating over time $t \in [0,\gamma]$, we have that
the total complexity is given by
\begin{equation}
N \int_0^\gamma \frac{\gamma-t}{1-t} k^2(t;\gamma) dt =
N^3 \int_0^\gamma (\gamma-t) (1-t) \rho^2(t;\gamma) dt \quad .
\end{equation}
Since the function $\rho(t;\gamma)$ is continuous in $t$, we conclude
that
\begin{eqnarray}
\left.
\begin{array}{c}
\rho(t;\gamma) \propto 1/N \\
k(t;\gamma) \text{ finite}
\end{array}
\text{, for all } t \in [0,\gamma] \right\} & \Leftrightarrow
\rho_{max}(\gamma)=0 \Leftrightarrow & \left\{
\begin{array}{c}
\text{CPU time} \propto N\\
\text{Memory} \propto N
\end{array}
\right.\\
\left.
\begin{array}{c}
\rho(t;\gamma) \text{ finite}\\
k(t;\gamma) \propto N
\end{array}
\text{, for some } t \in [0,\gamma] \right\} & \Leftrightarrow
\rho_{max}(\gamma)>0 \Leftrightarrow & \left\{
\begin{array}{c}
\text{CPU time} \propto N^3\\
\text{Memory} \propto N^2
\end{array}
\right.
\end{eqnarray}
where
\begin{equation}
\rho_{max}(\gamma) = \lim_{N\to\infty} \max_{t \in [0,\gamma]}
\rho(t;\gamma)
\end{equation}
is the order parameter signaling the onset of the hard regime.

Having found the relation between the density of ones in $L$ and the
computational complexity we are interested in, we can now run the
algorithms and measure the density $\rho(t;\gamma)$.  The easy/hard
transition should manifest itself with $\rho_{max}(\gamma)$ becoming
different from zero.

\subsection{Simplest Gaussian elimination}

Let us start with the simplest algorithm, which solves the equations
in the same (random) order they appear in the set and with respect to
a randomly chosen variable.  In this very simple case, one can easily
show that the complexity for solving a set of linear equations with
initial parameter $\gamma=\gamma_0$ is exactly the same as for solving
a larger system with $\gamma>\gamma_0$ up to time $t=\gamma_0$.  For
this reason, in this case the function $\rho(t;\gamma)$ does not
depend on $\gamma$ and can be calculated once for all the relevant
$\gamma$ values.

Moreover, it is known \cite{RIWEZE} that this algorithm, in the limit
of very large $N$, keeps the matrix sparse for all $\gamma < 2/3$.

\begin{figure}
\includegraphics[width=0.7\columnwidth]{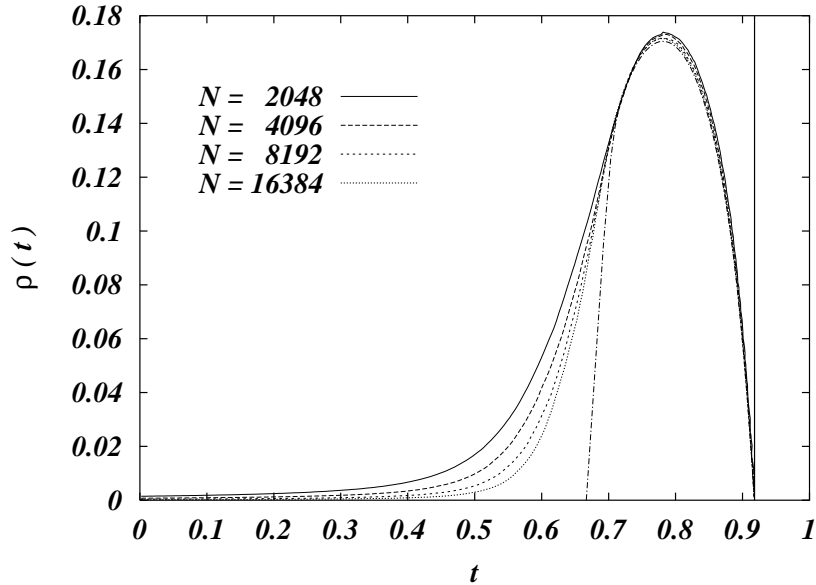}
\caption{Density of ones in the $L$ matrix during the solving process
with the simplest Gaussian elimination algorithm.  The vertical bar
marks the analytical critical point $\gamma_c=0.918$.}
\label{ge}
\end{figure}

In Fig.~\ref{ge} we show the function $\rho(t)$ for many large $N$
values.  The dotted-dashed line is a guide to the eyes and it should
not be too much different from the thermodynamical limit: It goes
through the two points ($\gamma=2/3$ and $\gamma=0.918$) where
$\rho(t)$ must vanish and coincide with numerical data in the region,
where data for different sizes seem to be quite close to the
asymptotic shape.

In the thermodynamical limit, the algorithm keeps the matrix sparse
for times $t \le 2/3$ and so it undergoes an easy/hard transition at
$\gamma = 2/3$: As long as $\gamma \le 2/3$, $\rho_{max}(\gamma)=0$,
while $\rho_{max}(\gamma)>0$ for $\gamma > 2/3$.  As we will see below
the location of the transition depends on the algorithm used and, in
this case, does not correspond to any underlying thermodynamical
transition.

We note \textit{en passant} that the $\gamma$ value where the $L$
matrix becomes sparse again seems to correspond to the critical point
$\gamma_c=0.918$ \cite{RIWEZE,MERIZE,CDMM} (marked with a vertical
line in Fig.~\ref{ge}).  An explanation to this observation will be
given in a forthcoming publication.  It implies that the value of the
critical point $\gamma_c$, which is relevant e.g.\ in the XORSAT model
\cite{XOR-SAT} in theoretical computer science, could be obtained also
by solving differential equations for $\rho(t)$.

\subsection{Smart Gaussian elimination}

Now we turn to a more clever Gaussian elimination algorithm, which
works as follows: At each time step, it chooses the variable $x$
having the smallest connectivity in $L$, i.e.\ that corresponding to
the less dense column of $L$, and solves with respect to $x$ any of
the equations where $x$ enters in.

Clearly, in this case, the dynamics and thus the density of ones in
$L$ depend on the initial $\gamma$ value: A smaller $\gamma$ implies
that for a longer time we can choose variables of connectivity 1,
which do not increase the average number of ones per row in $L$.  It
can be rigorously shown \cite{MERIZE,CDMM} that this procedure keeps
the density of the $L$ matrix constant, $\rho(t;\gamma) =
\rho(0;\gamma)$, for times smaller than $t^*=\gamma(1-m^3)$, where $m$
is the largest solution to $1-m=\exp(-3 \gamma m^2)$.  The last
equation is exactly Eq.~(\ref{mag}) with $\{a_3=1,a_{k \neq 3}=0\}$.

\begin{figure}
\includegraphics[width=0.7\columnwidth]{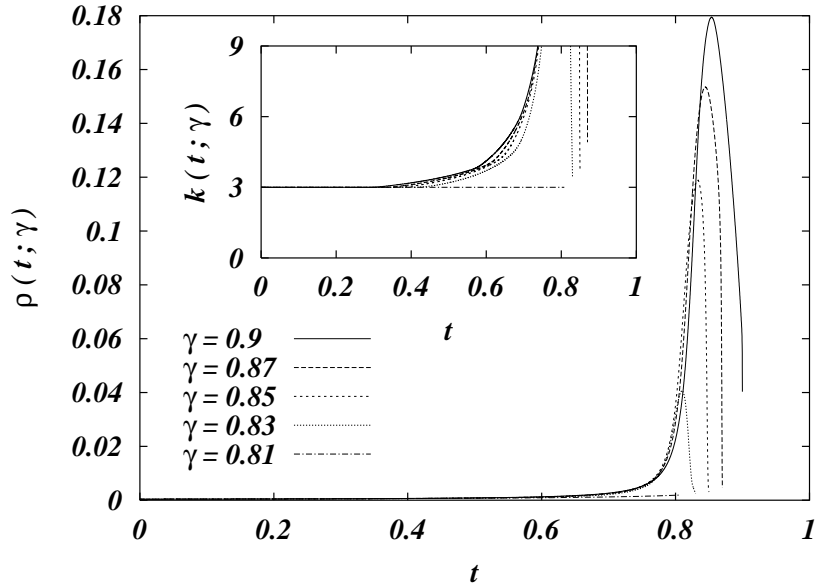}
\caption{Density of ones in the $L$ matrix during the solving process
with a smart Gaussian elimination algorithm ($N=8192$).  Inset: Zoom
on the low-density part (with a different normalization).}
\label{ges}
\end{figure}

Running the algorithm for different $\gamma$ values we obtain the
densities reported in the main panel of Fig.~\ref{ges}.  For
$\gamma<\gamma_d=0.818$ the density remains $\mathcal{O}(1/N)$ all
along the run, while for $\gamma>\gamma_d$ there is a time when the
density becomes finite and the problem hard to handle.

In order to better show what happens around $t^*$, we have plotted in
the inset of Fig.~\ref{ges} the mean number of ones per row, $k(t)$.
It is clear that for $\gamma<\gamma_d$ this number remains constant,
since one can solve the system choosing only variables of connectivity
1, not altering the $L$ matrix.  On the contrary, for $\gamma \ge
\gamma_d$ there is a time $t^*(\gamma)$ when variables of connectivity
1 terminate, and the algorithm has to start making substitutions in
$L$, thus increasing the density of ones.

Then $\gamma_d$ marks the onset of computational hardness, both in
memory and CPU time.  One may object that also this value for the
easy/hard transition may depend on the particular algorithm.  Note,
however, that a completely different linear algorithm described in
Ref.~\cite{RIWEZE} (which firstly works with high-connectivity
variables) seems to work up to $\gamma_d$.  Moreover, as seen in the
previous section, we have analytically found that at $\gamma_d$ a
transition takes place, which drastically changes the structure of the
solutions space, and so we argue that {\em any} algorithm running in
linear time can converge only up to $\gamma_d$.  Indeed is shown in
\cite{MERIZE} that solutions spontaneously form clusters for $\gamma >
\gamma_d$ and this particular structure requires a larger memory to be
stored.

\section{The RSA cryptosystem and factorization}

In this section we shall validate the above scenario on a concrete
application, namely integer factorization problems arising in the RSA
cryptosystem. Such problems allow for a non-trivial mapping onto huge
linear systems in GF(2) with a rather peculiar structure of the
underlying contact matrix.  In order to be as self-contained as
possible, we firstly give a short review of the problem and the
methodology (a detailed description of the RSA cryptosystem can be
found in \cite{RSA}).

The only known method for breaking RSA implies factorization of the
private key, which consists in a natural number which is the product
of two big prime numbers, $n=p\cdot q$, with $p$ and $q$ approximately
of the same size $\simeq \sqrt{n}$. Keys currently used in
applications are numbers $n$ ranging from 1024 bits (309 decimal
digits) to 2048 bits (617 digits) length.

The first attempt at a massive parallel factorization was the RSA129
(129 digits, 428 bits) challenge, solved in 1994 with the
\emph{quadratic sieve} (QS) algorithm. More recently, in August, 1999
the RSA155 challenged was solved using the \emph{general number field
sieve} (\emph{GNFS}) algorithm. This has forced to abandon the 512-bit
(155 digits) length for sensitive information security.

There are now several sub-exponential algorithms for solving the
factorization problem, the faster of which is \emph{GNFS. QS} and
\emph{GNFS} share the same structure, consisting of two phases: a
first one in which a big (the size depending mostly on the size of
$n$) linear system in GF(2) is produced, and a second one in which
this system is solved.

Although the first phase is definitely more costly, the solving phase
(which affect this paper) takes a respectable part of the total time
and memory requirement. Especially as numbers get bigger this becomes
a limitation, because the fastest solving methods used employ a sole
workstation, with the consequent memory restriction. Moreover, in
recent factorizations a new filtering phase has been placed between
the previous two, in which pieces of the system (specifically columns
of the $\{0,1\}-$matrix) get discarded in order to simplify the
solving phase, effectively transferring part of the total time from
the second phase to the first one.

In this section we will study statistical characteristics of linear
systems produced by the QS algorithm. Next subsection is dedicated to
a schematic description of QS.

\subsection{The QS algorithm }

For a nice description of the QS algorithm see \cite{CRYPTO}.
Synthetically, QS works at follows.  It builds a list of integer
numbers $\{y_{i}\}_{i\in I}$ such that:
\begin{itemize}
\item $y_{i}\equiv x_{i}^{2}\, (\text{mod}\, n)$ for some $x_{i}$ and
$y_{i}\neq x_{i}$;
\item $y_{i}$ is completely factorizable in a given (relatively small)
subset of $B$ primes called the factor-base.
\end{itemize}
This is called the \emph{sieving} phase.  The algorithm then searches
a subset $J\subset I$ of elements of the list such that $\prod _{i\in
J}y_{i}=z^{2}$ is a square (\emph{solving} phase). Once found,
$z^{2}\equiv x^{2}\, (\text{mod}\, n)$ (here $x=\prod _{i\in J}x_{i}$)
and this implies that $n$ divides $(x+z)(x-z)$ and then $gcd(x-z,n)$
will likely (further trials will increase the probability) be a
non-trivial factor of $n$.

\subsubsection{Sieving}

In order to find element pairs $x_{i}$,$y_{i}$ such that $y_{i} \equiv
x_{i}^{2} (\text{mod}\, n)$ we can use the polynomial $y=f(x)=x^{2}-n$
and evaluate it at different values of $x$, keeping only values of $y$
which completely factorize between the first $B$ primes (the
factor-base). The sieving will allow us to do this efficiently.

The idea is that, given $p$, it is easy to find which are the values
of $f(x)$ which are divisible by $p$, because $p$ divides $f(x)$ if
and only if $f(x)=x^{2}-n\equiv 0\, (\text{mod}\, p)$ and this is a
quadratic equation in GF(p), having at most 2 solutions.  These
solutions are nothing but the square roots of $n$ modulo $p$ (if they
exist).

This has a first consequence, i.e. that a prime $p$ will not divide
$f(x)$ if $n$ is not a square $\text{mod}\, p$ independently of the
value of $x$. So if we can detect these primes, we can eliminate them
directly from our set of primes.  Detecting them is very easy: Using
Fermat's little theorem, we know that
\begin{equation}
n^{p-1} \equiv 1 \, (\text{mod}\, p) \quad ,
\label{fermat}
\end{equation}
assuming that $p$ do not divide $n$ (which is trivially a reasonable
assumption, anyway, because we are searching a divisor of $n$).  If
$p$ is an odd prime, i.e. not 2 (all $n$ are a squares $(\text{mod}\,
2)$), then calling $m=n^{\frac{p-1}{2}}$ we have that $m^{2}\equiv 1\,
(\text{mod}\, p)$, so $m\equiv \pm 1\, (\text{mod}\, p)$.  This $m$
will prove to be handy.

If $n\equiv s^{2}(\text{mod}\, p)$ then $m\equiv s^{p-1}\equiv
1(\text{mod}\, p)$.  Conversely, if $m\equiv 1$ then $n$ is a square
modulo $p$ (not proven here). The number $m$ is called the Lagrange
symbol and can be computed efficiently in one of the firsts stages of
the algorithm.  Useful primes (those with $m=1$) are roughly a random
half of all the first $B$ primes. So now we will keep only this half
and redefine ``the first $B$ primes'' as ``the first $B$ primes with
$m=1$''.

Computing the square root modulo $p$ is a bit more difficult than
knowing that it exists, but can also be done efficiently.  For
instance, the easiest case is when $\frac{p+1}{4}$ is an integer, then
$\left(n^{\frac{p+1}{4}}\right)^{2}=n^{\frac{p+1}{2}}\equiv m n\,
(\text{mod}\, p)$.  As $m=1$ (or else there is no solution) then $\pm
n^{\frac{p+1}{4}}\, (\text{mod}\, p)$ are the required square roots.

Once we have computed the two solutions $f(x_{p}^{1,2})\equiv 0\,
(\text{mod}\, p)$, then adding $p,2p,3p,\dots $ to them we will obtain
\emph{all} $x$ such that $f(x)$ is divisible by $p$.

The sieve idea is to initialize an array with values of $f(x)$ for
consecutive $x\in [[\sqrt{n}],[\sqrt{n}]+M]$ indexed by $x$, and then
for each $p$ in our factor base to divide the corresponding arithmetic
progression of $\{f(x_{p}^{1,2}+kp),k=1,\dots \}$ by $p$.  At the end
those values which are completely factored between the primes in the
factor base will become $1$ (Well, not exactly. Some of them can have
multiple times the same prime factor. But we can set up a threshold
instead of $1$ below which we consider the number completely
factored. We can recheck afterwards). We take those values and put
their factorization in an array
\[
\begin{array}{ccc}
 & \begin{array}{ccc} f(x_{1}) & \cdots & f(x_{m}) \end{array} & \\
\begin{array}{c} p_{1}\\ \vdots\\ p_{B}\end{array}
 & \left[\begin{array}{ccc}
   \alpha _{1}^{(1)} & \cdots  & \alpha _{1}^{(m)}\\
   \vdots  & \ddots  & \vdots \\
   \alpha _{B}^{(1)} & \cdots  & \alpha _{B}^{(m)}\end{array}
   \right] & (\text{mod}\, 2)
\end{array}
\]

\subsubsection{Solving}

The \emph{solving} phase is conceptually simple: A solution of the
homogeneous linear system $\hat{A}v=0$ is a $\{0,1\}$vector $v$ which
represent correctly the subset $J$, in the sense that $v_{i}=1$if and
only if $i\in J$.

\subsection{The matrix ensemble}

\subsubsection{Correlations}

We have implemented the simplest QS described in \cite{CRYPTO} in
order to analyze the output matrix ensemble. We attempted to look for
correlations in the presence/absence of different primes in the set of
divisors of the variables $y_i$. Specifically we checked that there is
virtually no correlation between rows of the matrix: We have taken one
such output matrix (resulting from the factorization of a product of
two $20$ digits primes) and computed the covariance between the
corresponding spin variables $s_1,s_2$ of two rows $r_1,r_2$, the
averages being taken along different columns,
\[
\langle s_1 s_2 \rangle - \langle s_1 \rangle \langle s_2 \rangle
\quad .
\]
Once repeated for all $r_1<r_2$, we found that all pairs have
correlations in the interval $0 \pm 0.06$, a proportion of $0.9999$
pairs having correlations in $0 \pm 0.02$.

\subsubsection{Dependence on ``factorization hardness''}

We then examined dependence of the resulting distributions of ones per
row on the ``factorization hardness'' of the number $n$.  Typically
(depending on the algorithm) the complexity of factorization depends
on the size of the smallest prime divisor of $n$~\footnote{This is why
in RSA we choose $n=p\cdot q$ with $p,q\simeq \sqrt{n}$.}: For
instance, \emph{trial division} ends in exactly this amount of steps.
It was conjectured that this would be reflected in the structure of
the output matrix.

\begin{figure}
\includegraphics[width=0.7\columnwidth]{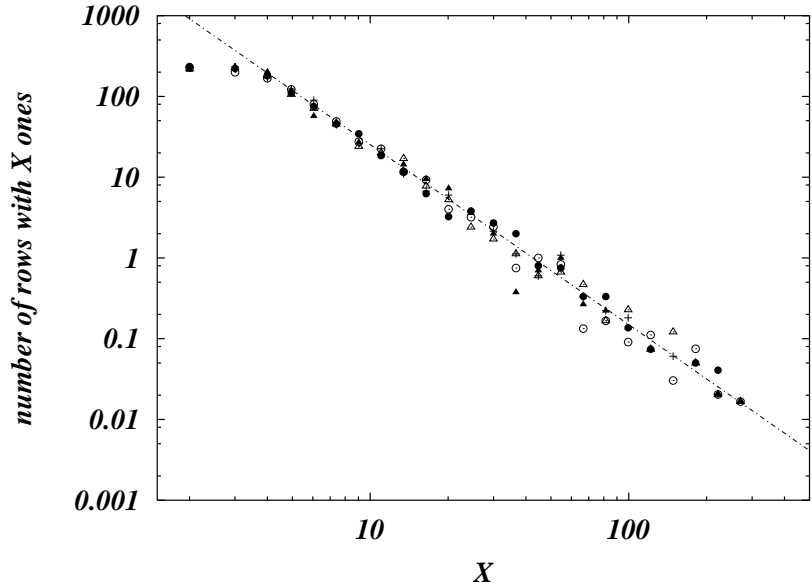}
\caption{Probability distribution of the number of ones per row in 5
different matrices of size around 1300.  The line is the best power
law fit on $X>3$ data, giving an exponent $\sim -2.2$.}
\label{realdata}
\end{figure}

We have constructed 25 numbers $n$ with different factor sizes (from
now on, factor type 10+10+10 will mean a 30 digit number constructed
as a product of tree 10-digit primes) organized as follows:
\begin{itemize}
\item 5 of type 20+20
\item 5 of type 10+30
\item 5 of type 13+13+13
\item 5 of type 10+10+10+10
\item 5 of type 5+5+5+5+5+5+5+5
\end{itemize}
All 25 numbers differed between them in less than a 0.01\%. We then
made QS compute the factorization matrices, with a factor base of size
1500.  This value for the size of the factor base has been chosen
experimentally in order to minimize the sieving phase duration.  The
resulting matrices were of size $1500 \times 1510$ and were then
post-processed in order to remove rows and columns with a single 1.
The final size is thus reduced of about 200 columns and rows.  The
resulting distributions of ones per row are plotted in
Fig.~\ref{realdata}, showing very little variations.  They can be very
well described by a unique distribution, which is substantially a
power law with some little deviations in the range of type-2 and
type-3 rows.  The best fit in the region $X>3$ gives an exponent $\sim
-2.2$.

Our conclusion is that statistical properties of the resulting matrix
do not depend on the factorization hardness.  The bottleneck for
factorizing a large hard number is mainly determined by the time
required by QS to generate the matrix, which indeed strongly depends
on the size of the smallest factor.  In the rest of the paper we will
analyze the solving phase, assuming the factorization matrix to have
uncorrelated rows and the number of ones per row to be a random
variable extracted from distribution in Fig.~\ref{realdata}.  These
two assumptions have been experimentally verified.

\subsection{Linear solving methods}

Plain standard Gaussian elimination execution time is cubic in the
size of the matrix (our matrices are almost square). Fortunately, we
can pack $32$ matrix entries in a single $4$-byte word, and then the
sum operation is implemented as the low-cost bit-wise logical XOR
operation, saving a factor 32 in time.

As also the matrix is very sparse, instead of keeping in memory all of
it, we can memorize only the position of 1's. This forbid us to use
the factor-32 trick, but allows us to do the first steps very
quickly. At some time in the Gaussian elimination process (typically
more than half of the process), the remaining (non eliminated) part
will be very dense, and then it will be convenient to switch to the
standard method above. This is what was done in the solving phase of
RSA129.

Another option is to use in one of the stages an iterative algorithms,
like the discrete Lanczos. The Lanczos method has the advantage of
having a stable $\mathcal{O}(N^{2})$ total time for a sparse matrix,
but finds only one solution (or a prefixed quantity in the
block-Lanczos variant) instead of all of them. For factorization this
is not a problem, because we need only a few solutions to have a
reasonable chance. This is the method that was used in the solving
phase of RSA155.

\subsection{Power law distributed $\{a_k\}$: Phase diagram and
comparison with real application data}

The previous analysis leads to the construction of matrices whose
density of non zero entries follows quite well a power law
distribution with light deviations due to rows with a small number of
ones and a cutoff, $k_{max}$, of some hundreds.  Then we use the
following distribution in the analytical treatment:
\begin{eqnarray}
a_2 &=& a \quad ,\\
a_k &=& \epsilon\; k^{-s} \quad \text{for\ \ } 3 \le k \le k_{max}
\quad ,
\label{eq:distr}
\end{eqnarray}
where $\epsilon$ is a normalizing factor equal to $(1-a) /
\sum_{k=3}^{k_{max}} k^{-s}$.  The factorized integers considered in
the previous section lead to an exponent $s \simeq 2.2$ and to a non
zero support up to $k_{max} \sim 200$.  The choice of keeping $a_2$,
and only $a_2$, as an independent parameter is dictated by the very
difference in the physical behavior of 2-spin terms and $k$-spin terms
with $k>2$.

The study of the phase diagram in the control parameter $\gamma$ for
choices of $a$, $s$ and $k_{max}$ retrieved from real data reveals a
non trivial behavior.

\begin{figure}
\includegraphics[width=0.8\columnwidth]{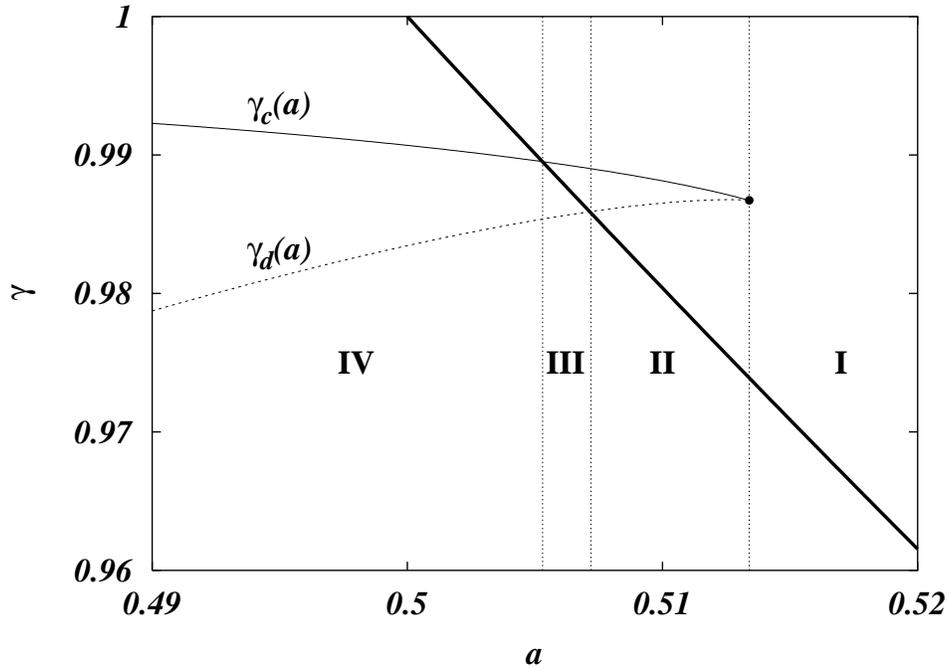}
\caption{Phase diagram ($a$, $\gamma$) for a typical choice of $s=2.2$
and $k_{max}=200$.  The bold line $1/(2a)$ represents a continuous
transition, while $\gamma_d(a)$ and $\gamma_c(a)$ corresponds
respectively to the spinodal and the critical lines of a first order
transition.  The dot marks the origin of these lines.}
\label{phase_diagr}
\end{figure}

\begin{figure}
\includegraphics[width=0.54\textwidth]{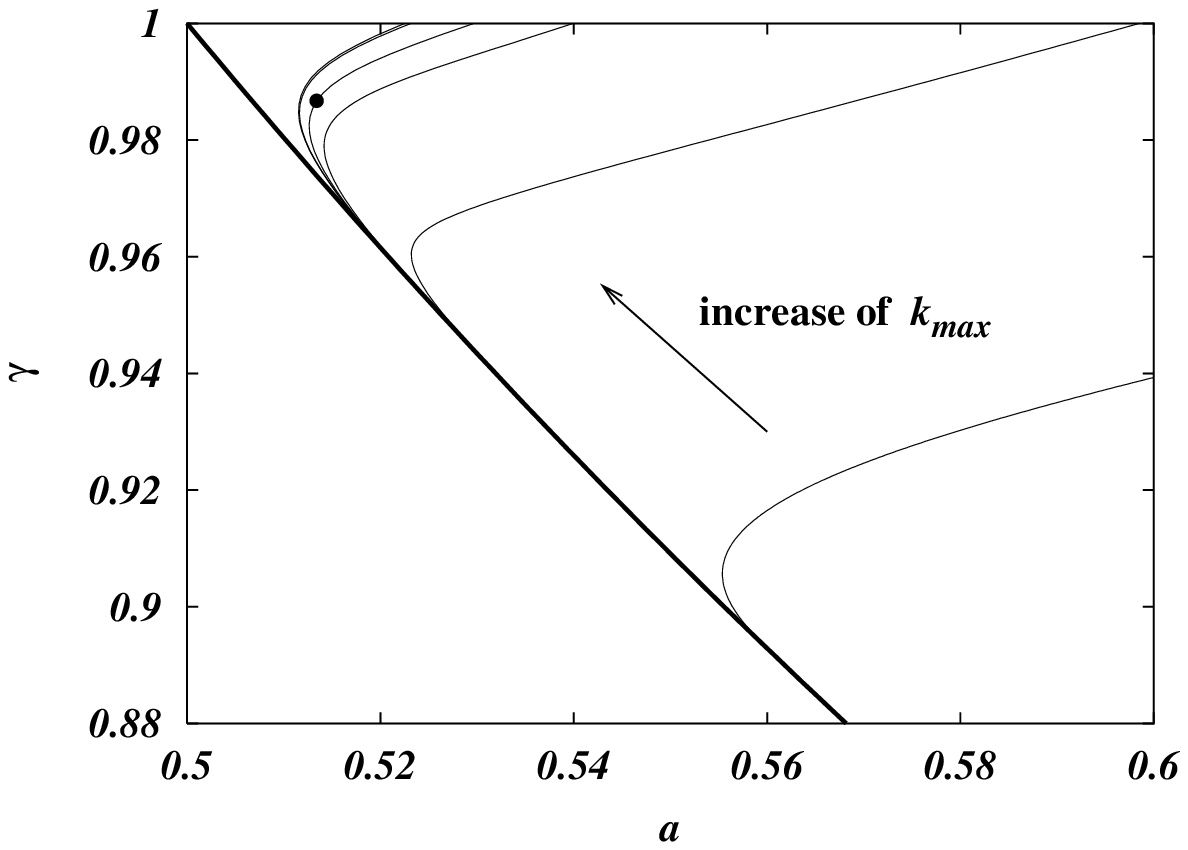}
\includegraphics[width=0.45\textwidth]{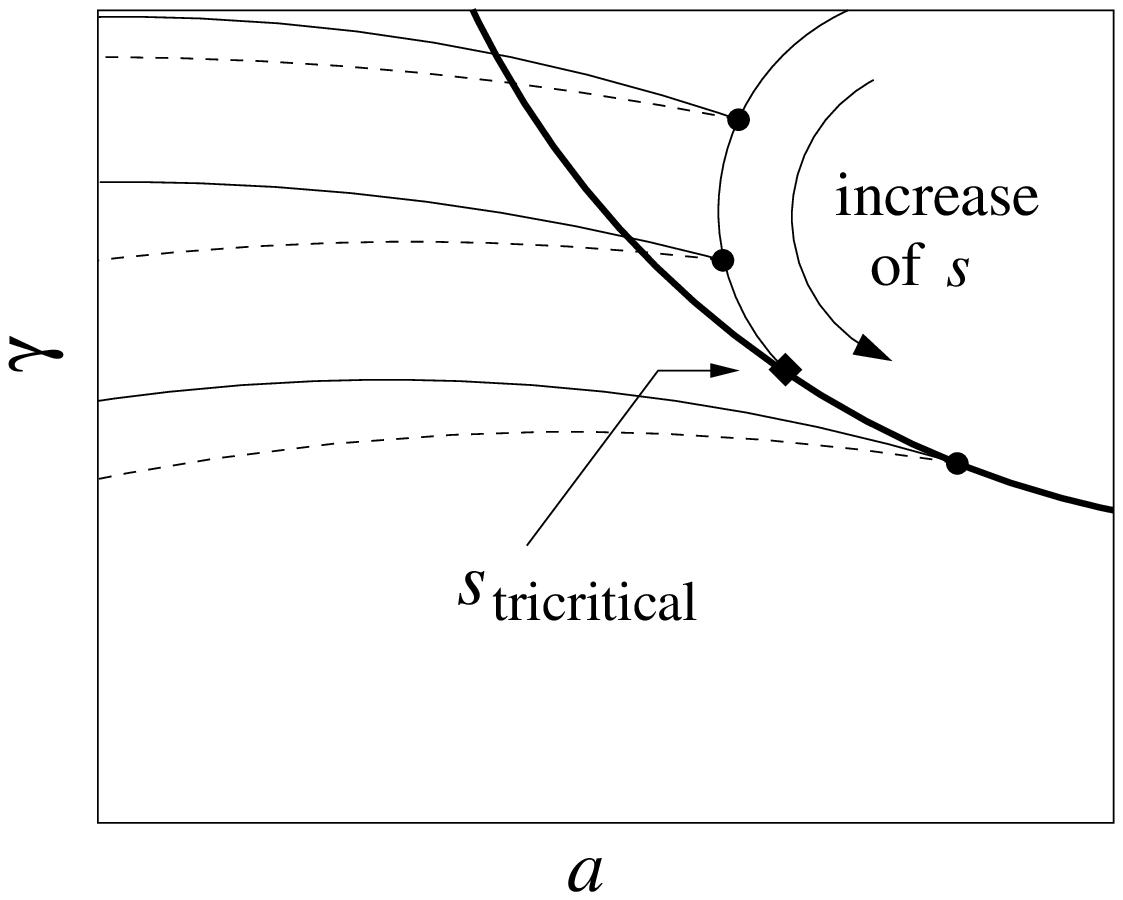}
\caption{Dependence on $s$ and $k_{max}$ of the origin of first order
critical lines.  The bold curve is the continuous phase transition
$\gamma = 1/(2 a)$.  Each solid bell-shaped curve in the left plot is
the ensemble of such origins, defined as the point where, decreasing
$a$, another non trivial solution to the saddle point equations
appears.  Each curve from right to left is indexed by a different
value of $k_{max} = 10, 30, 100, 200, 1000, 2000, 10000$.  Each point
on the curve corresponds to a particular value of $s$ (the dot is for
$s=2.2$ and $k_{max}=200$ as in Fig.~\ref{phase_diagr}).  Along the
curve $s$ increases for decreasing $\gamma$ (see right plot).  From
each point of the curve originate the two first order critical lines
shown for $s=2.2$ and $k_{max}=200$ in Fig.~\ref{phase_diagr}, and
pictorially drawn for different $s$ values in the right plot.  When
the origin joins the second order iperbole the system is at a
tricritical point.  $s_{\text{tricritical}}$ scales very rapidly with
$k_{max}$ converging to $\sim 2.73-2.74$ already for $k_{max} \sim
100$.}
\label{tricritical}
\end{figure}

In Fig.~\ref{phase_diagr} we show the phase diagram for $s = 2.2$ and
$k_{max} = 200$.  Only part of the entire phase diagram ($a \in
[0,1]$, $\gamma \in [0,1]$) is shown for clarity. The lines further go
on smoothly outside the drawn portion.

If $a$ is high enough, we are in the rightmost region {\bf I} of the
phase diagram, where algorithms smoothly find solutions to the system
and do not undergo any critical slowing down.  Indeed, crossing the
bold iperbole $\gamma = 1/(2 a)$ given by the condition
$\left.\partial G(m)/\partial m\right|_{m=0} = 0$, the system
undergoes a continuous transition in the order parameter $m$,
representing the fraction of variables taking the same value in all
the solutions.  The problem of finding solutions is always easy, as
for the case $\{a_2=1;a_{k \neq 2}=0\}$ explained before.

Decreasing $a$ we meet a first intermediate region {\bf II}, where the
birth of a meta-stable non-trivial saddle-point solution at $\gamma =
\gamma_d (a)$ is given by the solution of Eq.~(\ref{gammadin}).
However, algorithms should not be much affected by this meta-stable
state, because the system starts magnetizing continuously before,
crossing the bold line.  Increasing $\gamma$ up to the critical value
$\gamma_c(a)$ one meets a first order transition, where the
magnetization, that was already non-zero, undergoes a further jump.

The second central region {\bf III} shows an inversion between
$\gamma_d(a)$ and the bold line $1/(2a)$.  These two intermediate
regions have not been exhaustively studied yet, because real data all
fall in the leftmost one.  The shape of the central part of the phase
diagram is very sensitive to the choice of the control parameters $s$
and $k_{max}$, as shown in Fig.~\ref{tricritical}.

The $\gamma_c(a)$ curve in the second and third regions is found
solving
\begin{equation}
S(m^*,\gamma_c) = S(m_*,\gamma_c) \quad ,
\label{cross}
\end{equation}
where $m_*$ is the smallest positive solution to $G(m)=0$, which
corresponds to the magnetization of the ferromagnetic state arisen
from the second order transition (bold line).  The points of crossing
showing the onset of different regions, from right to left, are found
respectively as: $\frac{\partial G(m)}{\partial m} = 0\ \&\
S(m^*,\gamma) = S(m_*,\gamma)$,\ \ $\frac{\partial G(m)}{\partial m} =
0 \ \&\ \gamma = \frac{1}{2a}$\ \ and\ \ $S(m^*,\gamma) = S(0,\gamma)\
\&\ \gamma = \frac{1}{2 a}$.

The leftmost part {\bf IV} shows the typical behavior described in
\cite{RIWEZE}. Increasing $\gamma$ the system never reaches the
continuous transition on the bold line, but it undergoes a first
dynamical transition at $\gamma_d(a)$ and second thermodynamical one
at $\gamma_c(a)$, found via Eq.~(\ref{cross}) with $m_*=0$ since we
are still below the second order transition line.  Configurational
entropy is non-zero between $\gamma_d(a)$ and $\gamma_c(a)$, and
solving algorithms are affected by it.  There are typically other
spinodal lines in the phase diagram, but they always correspond to
sub-optimal solutions, and were, therefore, not shown in the picture.

In real data the fraction of 2-variables equations is typically of the
order of 0.2 and $\gamma \simeq 1$.  So we always work deep into phase
\textbf{IV} where, during the solving procedure, the system undergoes
a first dynamical transition, that corresponds to a slowing down of
the solving algorithms, before finding solutions.

\section{Conclusions}

We have analyzed the behaviour of different type of polynomial
algorithms in the solutions of large-scale linear systems over finite
fields.  The connection between memory requirements and clustering
phase transitions as been made clear on both artificially generated
problem as well as on a ``real-world'' applications.  While the role
of the dynamical glass transition in local search algorithm was
already well known (trapping in local minima), we have provided a
clear example of the role of such type of glass transition in global
dynamical processes which are guaranteed to converge to the global
optimum in some polynomial time. The memory catastrophe found is such
cases constitutes a concrete limitation for the performance of
single-machine programs.

\end{document}